\documentclass{pasj00}

\newcommand{\rosat}{{\it ROSAT}}

\newcommand{\chandra}{{\it Chandra}}
\usepackage{graphicx}
\usepackage[figuresright]{rotating}

\begin{document}
\SetRunningHead{Hudaverdi et al.}{Overdensity of X-ray Sources in the Field of Nearby Clusters of Galaxies}
\Received{2006/05/16}
\Accepted{2006/09/07}

\title{Overdensity of X-Ray Sources in The Field of Two Nearby Clusters of Galaxies: XMM-Newton View of A194 and A1060}

\author{Murat \textsc{Hudaverdi}, Hideyo \textsc{Kunieda}, Takeshi \textsc{Tanaka} and Yoshito \textsc{Haba}} 
\affil{Department of Physics, Nagoya University, Furo-cho, Chikusa-ku, Nagoya 464-8602, Japan}
\email{hudaverd@u.phys.nagoya-u.ac.jp}

\author{Akihiro \textsc{Furuzawa} and Yuzuru \textsc{Tawara}}  
\affil{EcoTopia Science Institute, Nagoya University, Furo-cho, Chikusa-ku, Nagoya 464-8602, Japan}

\and
\author{E. Nihal \textsc{Ercan}}
\affil{Bo\u{g}azi\c{c}i University, Physics Department, 80815 Bebek, Istanbul, Turkey}


\maketitle

\begin{abstract}
Two nearby clusters of galaxies: A194 ($z$$=$0.018) and A1060 ($z$$=$0.0114) have been 
analyzed for their X-ray point source properties with  \textit{XMM-Newton} EPIC-PN data.
A multi-band source detection technique was applied to both of the clusters, resulting 
in 46 sources from the A194 field and 32 sources from the A1060 field, respectively.	
The cumulative log($N$)-log($S$) for a flux limit of $F_X$$\geq$1$\times$10$^{-14}$ ergs cm$^{-2}$ s$^{-1}$
is calculated and compared with that of the Lockman Hole.
A $\sim$3$\sigma$ excess of X-ray sources is found for the cluster regions.
Considering the higher fraction observed in optical studies from the clusters, 
we estimate that the cluster source density is 6 times higher than the blank field source density,
and 15 times higher than the local group.  
Our X-ray selected sources have luminosity values between 10$^{39.6}$$\leq$$L_X$$\leq$10$^{41.4}$ ergs s$^{-1}$,
in which X-ray emission from LMXBs, hot halos and starburst galaxies becomes noticeable.
The significance of the source density excess vanishes gradually for sources with $L_X$$\geq$10$^{40.5}$ ergs s$^{-1}$, 
at which point the source density becomes comparable to that of the blank-field level.
Considering this confined low luminosity range and the X-ray to optical luminosity ratios ($L_X/L_B$),
the observed overdensity is ascribed to AGN fueling by its infall into cluster environment for 
$L_X$$\leq$10$^{40.5}$ ergs s$^{-1}$ in the X-ray luminosity function.
Whereas,  the quenching of AGN activity by the deep cluster potential explains 
why the excess of the source density vanishes for the brighter sources.
\end{abstract}

\section{Introduction}
X-ray studies of point sources within clusters are somewhat difficult,
because observations of the galaxies are normally impeded by luminous diffuse emission of 
the Intra-Cluster Medium (ICM) and limited angular resolution of the telescopes.
Recently, equipped with unprecedented spatial resolution and positional precision instruments, 
\textit{XMM-Newton} and ${\chandra}$ are uncovering higher fractions of galaxies in cluster fields
compared to non-cluster fields.

Optical surveys suggest that AGNs are relatively rare in clusters compared to the field (Osterbrock 1960).
Another survey by Gisler (1978) also found similar result based on a statictically improved sample.
Dressler et al. (1985 $\&$ 1999) report a fraction of $\sim$1\% AGN in the cluster environment, less than in the field (5\%).
Though, they also mention the possibility of a higher AGN fraction in nearby clusters.


   \begin{figure*}[htbp]
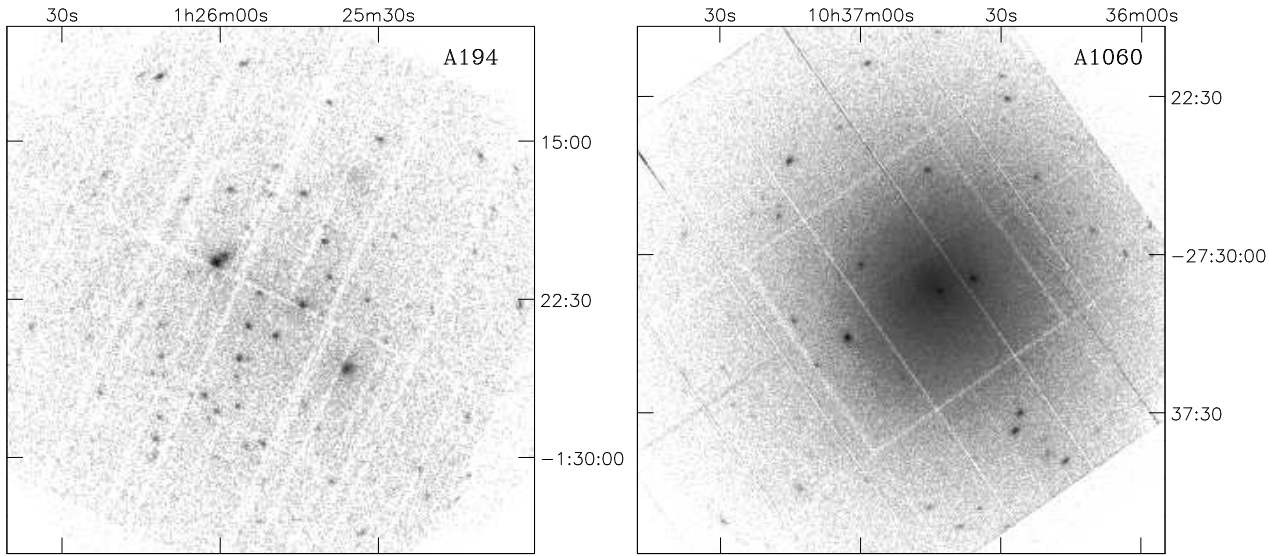

   {\centering
   \includegraphics[width=7.3cm,angle=270]{figure_1.ps}
   \includegraphics[width=7.3cm,angle=270]{figure_2.ps}
   }\caption{\label{images} 
Adaptively smoothed \textit{XMM-Newton} EPIC images of A194 (left) and A1060 (right) clusters in 0.3-10 keV band. 
Images have the size of 25$^{\prime}$ $\times$ 25$^{\prime}$ and pixel-size of 5$^{\prime\prime}$. 
The smoothing and contrast are selected in order to highlight point sources. 
The images are adjusted with north up and east left.}
    \end{figure*}


Several high-redshift clusters of galaxies have been reported to have an overdensity of X-ray point sources on their outskirts,
which is naturally attributed to infall related fueling of active nuclei.
Cappi et al. (2001) found 53 and 44 point sources from 3C295 and RX J0030 fields respectively, 
and reported an overdensity in these high redshifts ($z$ $\sim$ 0.5). 
Similarly, Molnar et al. (2001) detected significantly more sources around rich cluster Abell 1995 ($z=$ 0.32) than the fields.
They also surveyed MS0451 ($z=$ 0.55) but found no evidence of an excess, so concluded that 
different classes of objects seem to dominate in different clusters.
MS1054-0321 ($z=$ 0.83) was reported to have 47 X-ray sources by Johnson et al. (2003) and 
found to have a $\sim$2$\sigma$ excess compared to the non-cluster fields. 
Cappelluti et al. (2004) conducted a systematic analysis of the X-ray source population around 10 distant clusters (0.24 $\leq$ $z$ $\leq$ 1.2). They found excess values relative to the non-cluster fields more than a factor of 2$\sigma$ per field.
Ruderman $\&$ Ebeling (2005) analyzed ${\chandra}$ data of 51 clusters 
with redshift range 0.3 $\leq$ $z$ $\leq$ 0.7 and detected 8$\sigma$ 
significant excess in the point-source density within 3.5 Mpc of cluster centers.

At low redshifts, the excess seems to be largely due to low luminosity AGN associated with the cluster. 
One of the early study by Henry \& Briel (1991) with ${\rosat}$ for A2256 ($z=$ 0.06) 
found twice as many sources as expected from blank fields.
Lazzati et al. (1998), using the wavelet analysis technique on ${\rosat}$ PSPC observations of nearby clusters, 
A1367 ($z=$ 0.022) and A194 ($z=$ 0.018), detected 28 and 26 sources in the respective fields.
A1367 was also eventually observed by ${\chandra}$ and 59 sources are detected (Sun $\&$ Murray 2002).
Finally, Martini et al. (2006) spectroscopically confirmed a large AGN population in low-redshift clusters (0.05 $\leq$ $z$ $\leq$ 0.31).

In this paper, we report the X-ray point sources observed in the field of two nearby clusters
A194 ($z=$0.018) and A1060 ($z=$0.0114).
The emission from individual sources in these clusters stands out particularly well as the diffuse emission is not very luminous.
A194 is selected as a very faint ($L_{X}$ $\sim$ 10$^{42}$ ergs s$^{-1}$), 
poor cluster with no clear extended emission and no X-ray peak.
The cluster is known as linear cluster for its brightest member galaxies, which are aligned in the direction of northeast to southwest.
The temperature distribution of ICM is uniform at $kT$ = 2.7 keV. The cluster is also a strong radio emitter.  
Conversely, A1060 is X-ray brighter ($L_{X}$ $\sim$ 1 $\times$ 10$^{43}$ ergs s$^{-1}$) than A194.
The cluster is dominated by a pair of giant ellipticals, NGC 3309 and NGC 3311. 
A circularly symmetric smooth surface brightness, isothermal ICM ($kT$ = 3.3 keV)
and being populated by at least six bright NGC galaxies in the central region make the source very suitable for our survey.

We adopt  $H_o$=75 km s$^{-1}$ Mpc$^{-1}$ and cosmological deceleration parameter of $q_0$=0.5.
The luminosity distances to A194 and A1060 are 73 Mpc and 46 Mpc,
and an angular size of 1 arcmin corresponds to 20.5 kpc and 13 kpc, respectively.
The quoted uncertainties for the best fit parameters of spectral fittings are given for 90\% confidence range.

\section{Observations and data reduction}

We use X-ray data obtained by \textit{XMM-Newton} observations which cover a major portion of the cluster emission 
with large collecting area.
The three European Photon Imaging Camera (EPIC) instruments,
the two EPIC-MOS detectors (Turner et al. 2001) and the EPIC-PN detector (Str\"{u}der et al. 2001), are used.

The A194 cluster field was observed with the \textit{XMM-Newton} satellite in revolution 557 on December 2002.
The cameras were operated in Prime Full Window mode with the thin filters.
A1060 was observed on June 2004 in revolution 834. 
The medium filter was used for three EPIC cameras.
In our study, the Lockman Hole is chosen and referred to as the blank field to compare with the cluster fields.
We selected four recent long late observations of the Lockman Hole.
The total exposure is $\sim$400 ks.
All three cameras in the four observations were performed with the medium filter. 
The difference in the count rates from using medium and thin optical filters is negligible. 
It is less than 2\% at E $>$ 0.5 keV, and gets the highest value of $\sim$10\%  at 0.3 keV.
The summary of the observations is given in Table \ref{obs_log}.


\begin{table}[htb]  
\caption{\label{obs_log} Log of observations. (1) OBS-ID: observation id., (2) LH: Lockman Hole}
{\centering          
{\scriptsize
\begin{tabular}{l c c c c c l }     
\hline\hline       
Source    		&  OBS-ID$^{\mathrm{1}}$       & Date          & \multicolumn{3}{c}  {Live time (ks)} &  Filter  \\
         		&             	& 	        & MOS1 	& MOS2 	& PN 		&  EPIC         \\
\hline
A194   			& 0136340101 	& 2002.12.23  	& 18.6 	& 19.1	& 20.6		& thin   	\\
A1060			& 0206230101	& 2004.06.29	& 37.7	& 38.4	& 57.2		& med.	\\
LH$^{\mathrm{2}}$	& 0147510901 	& 2002.10.19	& 48.6	& 50.8	& 52.8		& med.	\\
				& 0147511601	& 2002.11.27	& 105.1 	& 106.6	& 97.2		& med.	\\
				& 0147511701	& 2002.12.04	& 92.7	& 93.4	& 86.2		& med.	\\
				& 0147511801	& 2002.12.06	& 77.5	& 78.7	& 63.9		& med.	\\

\hline                  
\end{tabular}}}
\end{table}


The event lists were generated from the Observation Data Files (ODF) by 
the tasks {\scriptsize EMCHAIN} and {\scriptsize EPCHAIN}.
We then selected the events with {\scriptsize PATTERN 0-12} for MOS and 
single $\&$ double pixel events ({\scriptsize PATTERN 0-4}) for PN.
In order to clean the contamination by soft proton flares, 
the extracted light curves for MOS (10-12 keV) and PN (12-14 keV) are 
clipped above and below 2.7$\sigma$ of the average count rate.
Finally, the resulting Good Time Intervals (GTI) selections are applied to the event lists to produce filtered event files.

\section{Source Detection}

We performed multi-band source detection on the EPIC images in
the soft band of 0.3$-$1 keV (S), medium band 1$-$1.6 keV (M), and hard band 1.6$-$10 keV (H).
The main concern with this energy selection is to identify diversity of sources with relative sensitivities at different energy bands.
Principally, the diagnosis of intrinsic absorption effects and power-law emission is intended.
This energy selection also splits the counts evenly which is statistically more favorable.

The source detection is performed with two different SAS tasks in order to carry out a comparable detection.
These are the maximum likelihood (Cash 1979) method {\scriptsize "EMLDETECT"} and 
the wavelet detection routine {\scriptsize "EWAVELET"} (Damiani et al. 1997).
The three X-ray images were scanned by using both methods.
The relation between likelihood limit ($L$) and gaussian probability ($P$) is given as $L = -\ln P $.
We selected a minimum likehood ML=10 and $4\sigma$ gaussian of signal-to-noise ratio,
at which a probability of $\sim$ $P = 3.2\times$10$^{-5}$ is attained by both values.
The methods take into account vignetting and PSF variations by means of the relative exposure maps.
The obtained lists are compared and sources falling outside of 25 arcmin circle are removed.
The final list is prepared by merging the three energy-band source lists for both PN and MOS
into a summary source list with the SAS command {\scriptsize "SRCMATCH"}.


\begin{table}[htb]  
 \caption{\label{source_numbers} Number of the detected sources for each camera and energy band. 
(1) Total number of sources for the final merged list., 
(2) LH: Lockman Hole,
(S) Soft: 0.3$-$1 keV, (M) Medium: 1$-$1.6 keV, (H) Hard: 1.6$-$10 keV,
(T) : Total number of sources detected by each camera.}
{\small
{\centering     
\begin{tabular}{l l l c c c c }     
\hline\hline       
Source    	& Total	&  	& \multicolumn{4}{c} {Source numbers}  \\
         	& number$^{\mathrm{1}}$	&       & S	& M	& H	& Total 	\\
\hline
A194   		& 56			&  MOS	& 30 	& 22	& 38	& 46  	\\
		         &			&  PN	& 37 	& 25	& 26	& 46		\\
A1060		& 32			&  MOS	& 20	& 17	& 17	& 27		\\
			&			&  PN	& 27	& 21	& 20 	& 32		\\
L.H.$^{\mathrm{2}}$	& 173	&  MOS	& 79	& 78	& 81	& 123	\\
			&			&  PN	& 98	& 81	& 54	& 123	\\
\hline                  
\end{tabular}}}
\end{table}

With the source detection routines mentioned above, 
the final merged lists contain 56 and 32 point sources for A194 and A1060, respectively.
The Lockman Hole observations have longer exposures and no contamination by extended 
X-ray emission as opposed to those of the cluster fields, 
thus the number of detected sources (173) is naturally higher. 
The sources are expected to have different spectral types, 
hence not all sources are detected significantly in all three energy bands.
It is also worth noting that some of the sources have been detected by only MOS or PN.
These sources are located at either in or around chip gaps. 
The source numbers for each camera and each energy band are given in Table \ref{source_numbers}.
The minimum flux values at 2$-$10 keV among the detected sources are 1.28${\times}10^{-15}$ erg cm$^{-2}$ s$^{-1}$ and 
1.23${\times}10^{-15}$ erg cm$^{-2}$ s$^{-1}$ for A194 and A1060.
We will use only PN data for the rest of the analysis,
because it has higher sensitivity than MOS cameras.

\section{X-ray color  diagrams}

We have developed a color diagram to determine average spectral properties of a given sample,
because not all of the sources have statistically sufficient photon counts to allow individual spectral fitting.
The color diagram is known to be a sensitive tool to define source types (White $\&$ Marshall 1984).
We defined two hardness ratio (HR) values; and 
plot soft ratio HR1=(M$-$S)/(M$+$S) against hard ratio HR2=(H$-$M)/(H$+$M). 
The hardness ratios are calculated from the net count rates in the soft, 
medium and hard bands obtained by the source detection routines.
The net counts of the three energy bands and hardness ratios for 
the detected sources are listed in the appendix (Table \ref{table_A194} and  \ref{table_A1060}).

   \begin{figure}
  { \centering
      \includegraphics[width=8cm,angle=0]{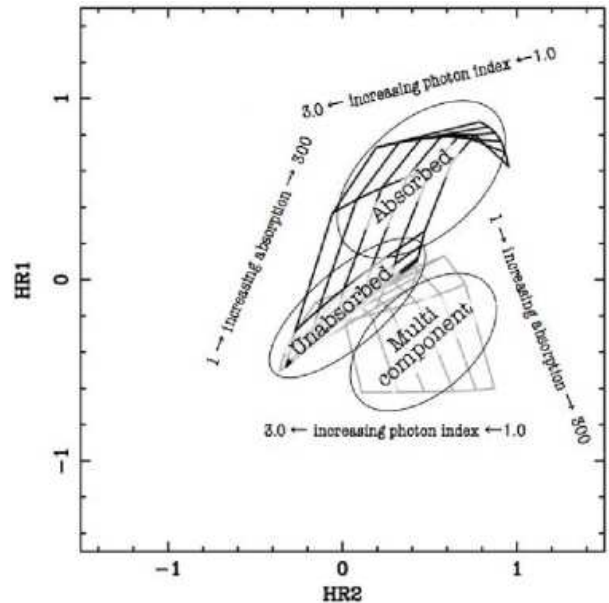}
     } \caption{ \label{simulation} The simulations of $zpha*(pow)$ (black) and $MEKAL+ zpha*(pow)$ (gray) are shown in grids.
X-ray colors of sources with power-law spectra with photon index 
$\Gamma$ gets the values from 1.0 to 3.0 in 0.1 steps as the grid moves left.
Increasing absorption ($N_{H}$=0.01, 0.05, 0.10, 0.50, 1.00 and 3.00$\times$10$^{22}$cm$^{-2}$) 
effects push sources toward the upper right in the diagram.}
            \end{figure}

%

   \begin{figure*}[htb]
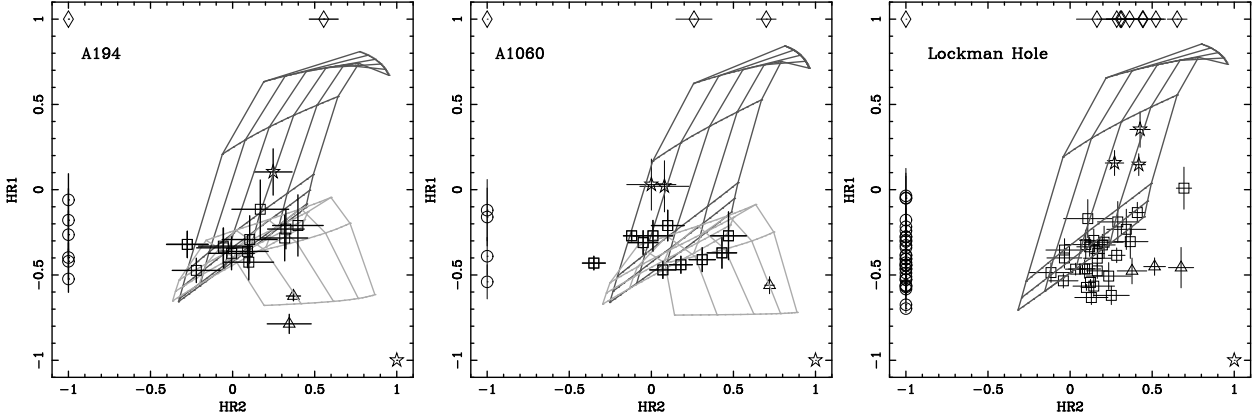

   \includegraphics[width=5.5cm,angle=270]{figure_4.ps}
   \includegraphics[width=5.5cm,angle=270]{figure_5.ps}
   \includegraphics[width=5.5cm,angle=270]{figure_6.ps}
   \caption{\label{color-diagrams} X-ray color diagrams of A194, A1060 and Lockman Hole 
showing the source classification defined in Table \ref{classification}. 
Each symbol represents different type of source;
Squares ($\Box$) are un-absorbed,  stars ($\star$) are absorbed, triangles ($\triangle$) are multi-component sources,
diamonds ($\Diamond$) are hard sources and circles ($\circ$) are soft sources.
The grids show the simulated models (see $\S$ 4.1 for details).}
    \end{figure*}

%

\subsection{\label{simu} Simulations}

We conducted simulations for each cluster to predict the locations of various sources in the color-diagram.
The simulations of both clusters are practically equivalent.
Therefore, we show only the grids from A194 simulations in Figure \ref{simulation}.
In all simulations, hydrogen column densities are fixed at galactic absorption values;
3.8$\times$10$^{20}$cm$^{-2}$, 4.9$\times$10$^{20}$cm$^{-2}$ and 
5.7$\times$10$^{19}$cm$^{-2}$ for A194, A1060 and the Lockman Hole, respectively.
We used a fixed metal abundance of 0.3 of solar in our simulations, which is the average galactic value.
The black grid shows the variation of power-law model with intrinsic absorption ($zpha*pow$ of {\scriptsize XSPEC}).
As X-ray colors of the sources move to the left (HR2 decreases) on the grid, 
the photon index $\Gamma$ increase from 1.0 to 3.0 in 0.1 steps.
The increasing absorption ($N_{\rm H}$=0.01, 0.05, 0.10, 0.50, 1.00 and 3.00$\times$10$^{22}$cm$^{-2}$) 
pushes the sources toward the upper right. 
In other words, as the absorption increases both HR1 and HR2 values increase.
We also simulated a multi-component model of {\scriptsize MEKAL} (Mewe et al. 1985) 
and absorbed power law for the galaxies with thermal plasma emission.
The power-law parameters are retained at the same values, whereas a temperature value of 0.5 keV is used.
This value is sufficient to represent sub-keV levels of thermal emission as produced by hot gas of normal galaxies.
The black grid of simple power-law model is entirely inclined down with the additional thermal model.
This curved grid is colored in gray (Figure \ref{simulation}) and 
displays the expected location of multi-component sources in the color-diagram.

\section{Classification of sources}

In this section, we explain the type of sources based on the definitions from the diagram.
A uniform classification criteria is defined for the two clusters.
Table \ref{classification} shows our classification of sources based on HR values of the X-ray color diagrams.
The classification scheme is roughly sketched with the ellipses on the grids of Figure \ref{simulation}.
We acknowledge that the color diagram is admittedly basic,
but it can be regarded as a reasonable method to infer appreciable knowledge on the physical nature of faint sources.

\begin{table}[htb]           
{\small      
\caption{\label{classification}X-ray color classifications. 
$(a)$ Sources with no detectable medium-band emission,
$(b)$ Sources with no hard-band emission, 
$(c)$ Sources with no soft-band emission,
$(d)$ Sources with only medium-band emission. 
The sources with only-soft or only-hard band detectable emission are classified 
as indeterminate sources and do not appear in this table. 
Their hardness ratio values are, as they get the name after, $indeterminate$.}
\begin{tabular}{l l  }     
\hline\hline       
            Classification      &  Definition \\
            \noalign{\smallskip}
            \hline
            \noalign{\smallskip}
            Unabsorbed  		&	-0.6 $\leq$ HR1 $\leq$ 0.0, -0.4 $\leq$ HR2 $\leq$ 0.5 	     	\\
            Absorbed  			&	 0.0 $\leq$ HR1 $\leq$ 0.8;	(-1,1)$^{\mathrm{(a)}}$ 		\\
            Multi component		&	 HR1 $\leq$ -0.4, 0.3 $\leq$ HR2 		     		\\
            Super Soft			&	 HR2 = -1.0$^{\mathrm{(b)}}$    \\
            Super Hard			&	 HR1 = 1.0$^{\mathrm{(c)}}$; (1,-1)$^{\mathrm{(d)}}$    \\
            \noalign{\smallskip}
\hline                  
\end{tabular}}

\end{table}
 
\subsection{Unabsorbed sources}

Most of the sources, which are detected in all three-bands, are fitted with single power-law model.
The photon indexes are in the physically tolerable range (0.8 $\leq$ $\Gamma$ $\leq$ 2.5), 
with the majority of sources clustering at around $\Gamma$=1.7 value. 
The sources with hardness ratio value of $-$0.6 $\leq$ HR1 $\leq$ 0.0, $-$0.4 $\leq$ HR2 $\leq$ 0.5 
are defined as unabsorbed sources.
We found 13 and 11 sources falling into this class for A194 and A1060, respectively.
The source \#4 from A1060 field is a good example of this kind.
Figure \ref{source_spectra}-A shows the single power law model 
with a photon index $\Gamma$=2.03$^{+0.27}_{-0.24}$.
The source has 750 counts in the total band 0.3$-$10 keV and 
the luminosity is $L_X$=10$^{40.2}$ ergs s$^{-1}$ in 2$-$10 keV band. 

\subsection{Absorbed sources}

\begin{figure*}[htbp]
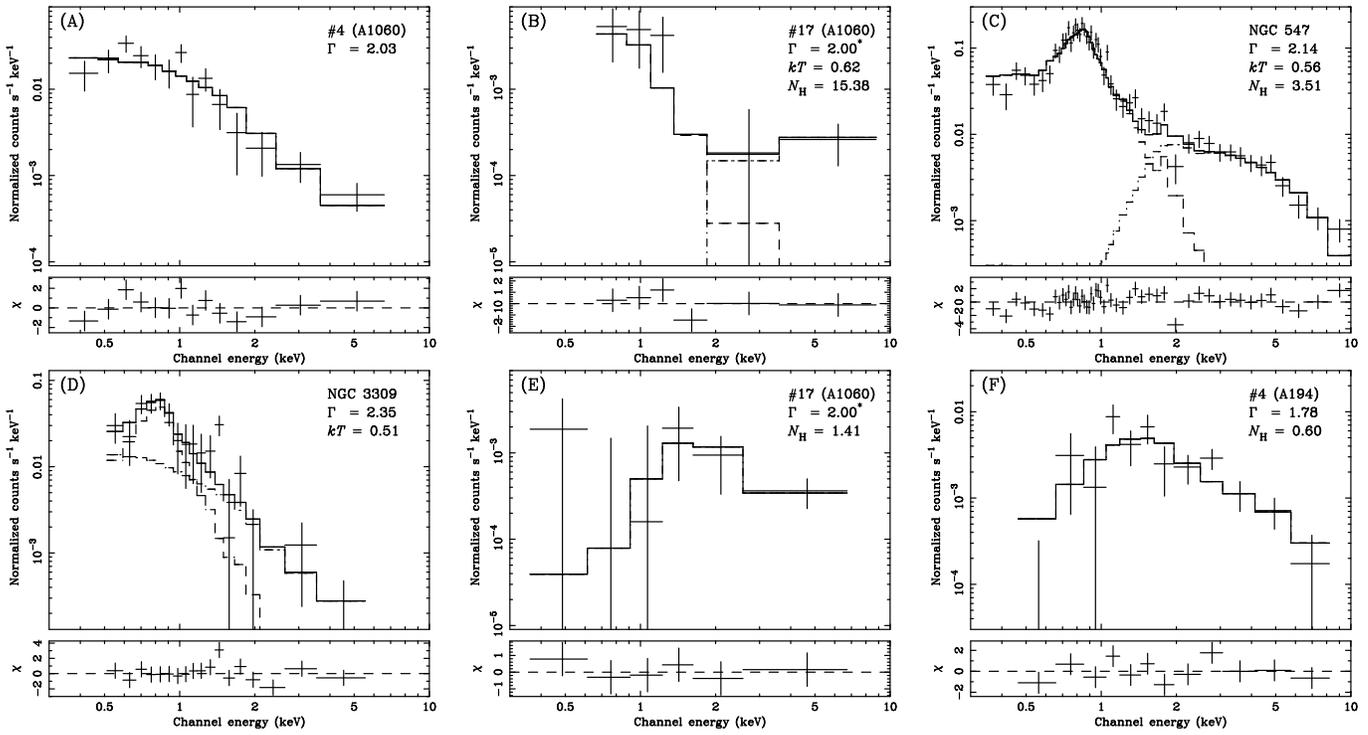

   {\centering
   \includegraphics[width=4.8cm,angle=270]{figure_7.ps}
   \hfill
   \includegraphics[width=4.8cm,angle=270]{figure_8.ps}
   \hfill
   \includegraphics[width=4.8cm,angle=270]{figure_9.ps}
   \hfill
   \includegraphics[width=4.8cm,angle=270]{figure_10.ps}
    \hfill
   \includegraphics[width=4.8cm,angle=270]{figure_11.ps}
    \hfill
   \includegraphics[width=4.8cm,angle=270]{figure_12.ps}
   }\caption{\label{source_spectra} 
EPIC-PN spectra with best-fit models. A: source \#4 (A1060), B: source \#17 (A1060), C: source \#1 - NGC 547 (A194),
D: source \#5 - NGC 3309 (A1060), E: source \#14 (A1060), F: source \#4 (A194).}
    \end{figure*}

The simulations showed (Section $\S$\ref{simu}) 
that increasing absorption pushes the sources toward the upper right in the diagram. 
In Figure \ref{color-diagrams} we highlighted the sources with the soft range 0.0 $\leq$ HR1 $\leq$ 0.8 
by star signs in the diagrams.
These are defined as the absorbed sources.
The sources are well fitted with absorbed power-law model.
The intrinsic absorptions values are $N_{\rm H}$ $\geq$ 0.5$\times$10$^{22}$ cm$^{-2}$.
The source \#17 from A1060 is one of the brightest sources in this classification.
About 60\% of the emission is detected below 1 keV and the remaining 40\% is from above 1.6 keV, 
with no detectable emission from 1$-$1.6 keV medium band.
Figure \ref{source_spectra}-B shows the X-ray spectrum of the source.
Best fit parameters are $N_{\rm H}$=15.38$\times$10$^{22}$ cm$^{-2}$, 
$\Gamma$=2.00 (fixed) and a thermal component of $kT$=0.62$^{+0.97}_{-0.39}$ keV.
The rest frame luminosity is $L_X$=10$^{39.5}$ ergs s$^{-1}$ for PN (120 counts in 0.3$-$10 keV band).
Absorbed sources with no detectable medium-band emissions are seen at HR1, HR2 = (-1, 1) point on the color diagrams.
4 sources from A194 and 6 sources from A1060 are classified in this group (Table \ref{population}).

\subsection{Multi component sources}

The spectra of some sources are not acceptably fitted with a single power law, the presence of a soft excess is evident.
The fit is significantly improved by adding a thermal plasma model to the main power law continuum.
The thermal model is parameterized by {\scriptsize MEKAL} plasma model (Mewe et al. 1985).
The hardness ratio values between HR1$\leq$$-$0.4 and HR2$\geq$0.3 correspond to multi-component sources.
We detected 3 sources in the cluster fields, 2 of which in the A194, and 1 in the A1060 field. 
The spectra of the source \#1 - NGC 547 (Figure \ref{source_spectra}-C) from A194 and \#5 - NGC 3309 
(Figure \ref{source_spectra}-D) from A1060 display the characteristics of this group.
NGC 547 is classified as type-2 AGN at $z$=0.01823.
The best-fit parameters are $kT$=0.56$\pm 0.03$ keV, $\Gamma=2.14 ^{+0.82}_{-0.73}$ and 
$N_{\rm H}= 3.51^{+2.62}_{-1.98}\times10^{22}$ cm$^{-2}$.
The source has rest-frame luminosity of $L_X$=10$^{42.1}$ ergs s$^{-1}$ in the 2$-$10 keV band.
The photon count is 1375 in 0.3$-$10 keV band.
NGC 3309 has a thermal gas with a temperature of $kT=0.51^{+0.13}_{-0.18}$ keV,
and a power-law index of $\Gamma=2.35^{+0.42}_{-0.77}$. 
The source has a luminosity of $L_X$=10$^{39.8}$ ergs s$^{-1}$ in 2$-$10 keV band.                                                                                  
Spectral best-fit temperature values of the multi-component sources are confined to 0.5 $\leq$ $kT$ $\leq$ 1.0 keV,
which is the anticipated range for thermal emission from normal galaxies.

\subsection{Super Soft \& Hard Sources}

We have detected 8 sources with no significant emission above 1.6 keV.
These sources which locate at HR1=$-$1 in the color diagram 
are defined as super soft sources (SSS) (except HR2, HR1=(1,$-$1), see $\S$ \ref{indeterminate}).
The features of these sources are quite similar to those found in Large Magellanic Cloud and M31 
(e.g., Kahabka \& van den Heuvel 1997; Pietsch W. et al 2005). 
On the contrary, 7 sources have no detectable emission below 1.0 keV, which are consequently named as super hard sources (SHS).
The diamond marks in the color diagrams (Figure \ref{color-diagrams}) stand for SHS.
As stated by the definitions, SHS get the value of HR1=1.0 and they are found at the top of the diagrams.
The sources \#14 of A1060 and \#4 of A194 exhibit the characteristic of SHS. 
The counts of these sources in the 0.3$-$10 keV band are 104 and 149, respectively.
Figure \ref{source_spectra}-E and Figure \ref{source_spectra}-F show the spectral fits with a simple power law model.
The photon indexes of sources \#14 and \#4 are $\Gamma$=2.00 (fixed) and $\Gamma$=1.78$^{+0.82}_{-0.65}$.
The intrinsic absorption values appear to be marginal but significant relative to the galactic values.
The best-fit values are $N_{\rm H}$=$1.41^{+5.88}_{-1.08}\times10^{22}$ cm$^{-2}$ and 
$N_{\rm H}$=$0.60^{+0.83}_{-0.37}\times10^{22}$ cm$^{-2}$, respectively.
The rest frame luminosities are $L_X$=10$^{40.1}$ ergs s$^{-1}$ and $L_X$=10$^{40.7}$ ergs s$^{-1}$ in the 2$-$10 keV band.


   \begin{figure*}
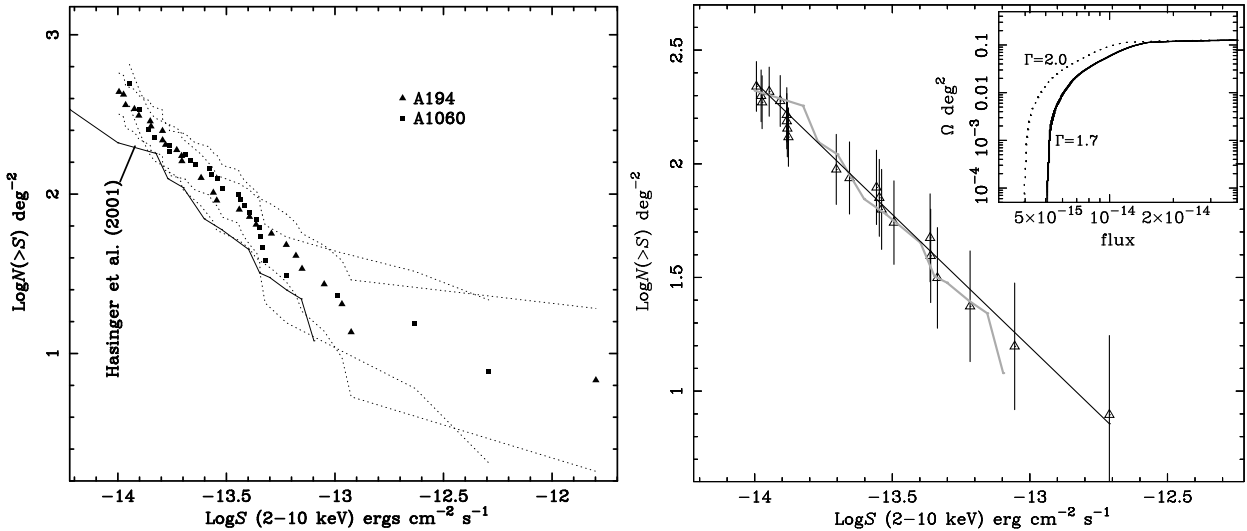

   {\centering
      \includegraphics[width=7cm,angle=270]{figure_14.ps}
      \includegraphics[width=7cm,angle=270]{figure_15.ps}
      }\caption{ \label{logN-logS} 
$\emph{Left panel:}$ log($N$)-log($S$) of the identified sources of our samples 
A194 (triangle - $\triangle$) and A1060 (box - $\Box$) is compared with 
the field of the Lockman Hole (Hasinger et al. 2001). The dotted lines show the statistical and calibration errors.
$\emph{Right panel:}$ log($N$)-log($S$) for Lockman Hole is fitted with a power-law model $N (> S) = K \times S^{-\alpha}$. 
The best fit values are ${\alpha} =1.16_{-0.25}^{+0.20}$ and $K=1.45{\times}10^{-14}$ deg$^{-2}$.
The model and the Lockman Hole plot of Hasinger et al. (2001) are shown in black and gray lines, respectively.
Error bars represent 1$\sigma$ statistical constraints and 10$\%$ absolute calibration errors (Kirsch 2006).
Survey area, $\Omega$ deg$^{2}$, is plotted against the sensitivity flux limit at the top corner. 
The solid and dashed lines correspond to the photon index values of $\Gamma$ = 1.7 and $\Gamma$ = 2.0 to convert the count rate to flux.
}
\end{figure*}

\subsection{\label{indeterminate} Indeterminate Soft \& Hard Sources}

A further classification of the sources is given for the indeterminate sources.
These sources have detectable emission only below 1 keV or only above 1.6 keV making their hardness ratio values undefined.
They can not be classified by a criteria based on the color diagram,
thus we name them as indeterminate sources.
The indeterminate source numbers in the cluster field are 18 for soft sources and 7 for hard,
while 31 for soft and 10 for hard sources in the Lockman Hole field.
The majority of them is on the outskirts (R $\geq$ 5 arcmin) and might be unrelated to the clusters.
The nature of these sources may include several types and are open to debate. 

The number distribution of the sources for the classification mentioned above is given in Table \ref{population}.
The number of soft sources in the Lockman Hole is exceptionally high, about half of the sources.
This is mainly due to higher efficiency of the detector at lower energies and 
no extended ICM emission in the Lockman Hole field that would bury point-like emissions.
In our study, The Lockman Hole is used as non-cluster control field to compare with the cluster regions,
therefore the classification in Table \ref{population} is sufficient for our objectives.
A further separation is beyond of the scope of our paper. 
The reader is referred to review of Mainieri et al. (2002) 
for a detailed discussion of the physical nature of the sources found in the Lockman Hole 
combining X-ray data with the optical/near IR data.



\begin{table}
\label{table:1}      
\caption{\label{population} The number of sources in each group.}             
\centering          
\begin{tabular}{l c c c }     
\hline     
Type    		& A194	& A1060 &  Lockman.H. \\
\hline                  
Unabsorbed 		& 13	& 11	& 28	\\
Absorbed		& 4	& 6	& 4	\\
Multi Component		& 2	& 1	& 3	\\
Super Soft		& 5	& 4	& 32	\\
Super Hard		& 3	& 4	& 15	\\
Indeterminate Soft	& 12	& 6 	& 31	\\
Indeterminate Hard	& 6	& 1 	& 10	\\
\hline                  
TOTAL			& 46	& 32	& 123	\\
\end{tabular}
\end{table}


\section{\label{lognlogs}  log($N$)-log($S$)}

The sum of the inverse areas of all sources brighter than flux $S$, 
$N(>S)$, is defined as the cumulative number per square degree.
Using the integrated formula of source numbers in units of deg$^{-2}$:
\begin{equation}
N (> S) = \sum_{i=1}^{n} \frac{1}{\Omega_i} ($deg$^{-2})
\end{equation}
where $n$ is the detected source number and $\Omega_i$ is the sky coverage for
the flux of the $i$-th source, the relation between number and flux is derived.
Figure \ref{logN-logS} left panel shows comparison of the obtained log($N$)-log($S$) plot 
for A194 \& A1060 and that of the Lockman Hole measured by Hasinger et al. (2001). 
The dotted lines represent the sum of 1$\sigma$ poisson errors in source counts and the systematic instrumental errors.
The systematic instrumental errors related to vignetting and point-spread function (PSF) 
are considered to be 3.5$\%$ for an off-axis source by Saxton (2003).
We use 10$\%$ as absolute calibration error, which dominates the other sources of error (Kirsch, 2006).
The limiting flux levels are 1.284$\times$10$^{-15}$ ergs cm$^{-2}$ s$^{-1}$ (A194) and
1.279$\times$10$^{-15}$ ergs cm$^{-2}$ s$^{-1}$ (A1060) in 2$-$10 keV band.
The characteristic of the solid angle is the same for the clusters and the field: the survey area decreases with flux.
In clusters, due to extended emission, faint sources can not be detected easily in the center, thus they rise at the outskirts.
Whereas, in the field there is no extended emission to bury faint sources.
Faint sources are detectable only at smaller off-axis angles where PSF and vignetting effects are smaller. 
We used a flux limit of 1$\times$10$^{-14}$ ergs cm$^{-2}$ s$^{-1}$ in our survey to avoid this ambiguity.
A significant ($\sim$3$\sigma$) X-ray source excess is found from cluster fields relative to the Lockman Hole.

In Figure \ref{logN-logS} right panel we present our Lockman Hole source numbers at 2$-$10 keV
which are in excellent agreement with the result of Hasinger et al. (2001).
The Lockman Hole log($N$)-log($S$) is described by a power-law model as a function of dimensionless $S$:
\begin{equation}
N (> S) = 1.45 {\times} 10^{-14} \times S^{-1.16 _{-0.25}^{+0.20}}
\end{equation}
with the slope of ${\alpha} =$ 1.16 $_{-0.25}^{+0.20}$ and the normalization of $K=1.45{\times}10^{-14}$ deg$^{-2}$.
The solid line shows the best-fit model (Figure \ref{logN-logS} right panel) comparing with the result of Hasinger et al. (gray line).
The survey area, $\Omega$ deg$^{2}$, as a function of limiting fluxes in the 2$-$10 keV band can be seen in the same figure as well.
Energy conversion factor (ECF) to derive flux from counting rate is calculated as 0.51${\pm}$0.09$\times$10$^{-11}$ ergs cm$^{-2}$ s$^{-1}$,
with the assumptions of $\Gamma$=1.7 and galactic value of hydrogen column density. 
The ECF errors correspond to photon index range of $\Gamma$=1.7${\pm}$0.3.

Log($N$)-Log($S$) conventionally shows source numbers per unit sky area.
An optimum comparison of the clusters with the field would be source number density per unit volume, 
as outlined by Finoguenov et al. (2004).
Unfortunately, in our survey distance information is not accessible for all the sources.
The photometric redshift value is available only for $\sim$10\% of the X-ray detected sources.
We use a subtraction method for an implicit approximation, 
assuming that the source density enhancement is attributable to 
the density contribution of galaxies in the cluster gravitational potential.
The method deduces the Lockman Hole source density from that of the clusters by carefully dealing with the errors.
It is a tricky job because of the statistical errors and cosmic variance which is field to field variation of the blank field.
Based on the most sensitive measurement to date,
 the cosmic variance is reported to be less than 15\% in the 2$-$10 keV band (Cappelluti et al. 2005).
Figure \ref{logN-logS} shows that our source density values correspond to 
438$\pm$87 sources deg$^{-2}$ for A194 and 
494$\pm$97 sources deg$^{-2}$ for A1060 around our flux limit of 1$\times$10$^{-14}$ ergs cm$^{-2}$ s$^{-1}$.
At this flux level, a density of 220$\pm$33 sources deg$^{-2}$ from Lockman Hole is calculated (errors, 15\% cosmic variance). 

\begin{figure}
   {\centering
      \includegraphics[width=6.5cm,angle=270]{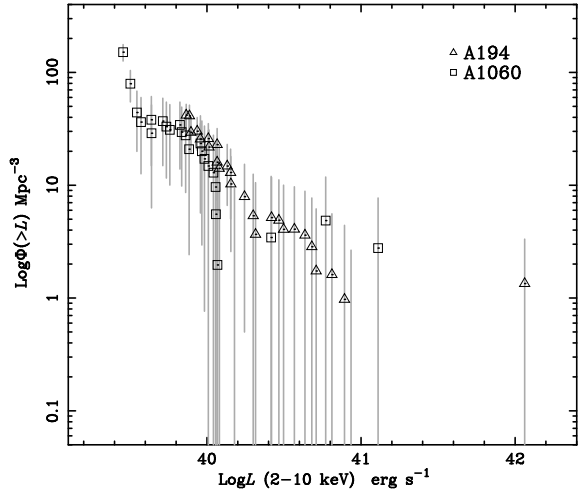}
    }
      \caption{\label{volume_density}
Integrated source density $\Phi$ against $L_X$ for 
A194 (triangle - $\triangle$) and A1060 (box - $\Box$).
There is a confusion around high luminosity levels since the number density of the sources at 
these luminosity levels is close to the density of the Lockman Hole (See $\S$ \ref{lognlogs}).
   }\end{figure}

If we perform the calculations by using the minimum possible source density values 
from clusters and maximum source density level for the Lockman Hole,
the number densities are 130  sources deg$^{-2}$ for A194 and 144  sources deg$^{-2}$  for A1060. 
This calculation for the worst-case scenario verifies that at least $\sim$30\% of the detected X-ray sources are cluster members.
Encouraged by this result, we estimate the unbinned X-ray Luminosity Function (XLF) 
per unit volume $\Phi(L_X)$ (Mpc$^{-3}$) as shown in Figure \ref{volume_density}.
The function is calculated from the relation:
\begin{equation}
  \Phi(L) = \frac {N}{\int_{L_{min}}^{L_{max}} \int_{z_{min}(L)}^{z_{max}(L)},\Omega(L,z)\,(dV/dz)\,dz\,dL},
\label{xlf_equ}
\end{equation}
where $N$ is the number of objects with luminosities and redshifts in 
the interval of $\Delta{L}\Delta{z}$ and $dV/dz$ is the volume element.
The method is described in detail by Page \& Carrera (2000) and Georgantopoulos et al. (2005).
The plot covers a luminosity range of 39.5 $\leq$ log($L_X$) $\leq$ 42.1 ergs s$^{-1}$.
There are a few luminosity levels with negative slope in the plot.
These luminosity levels are the points where the source density of the clusters is comparable to that of the Lockman Hole.

\section{\label{disc} Discussion}

We have analyzed \textit{XMM-Newton} observations of clusters of galaxies A194 and A1060.
We have derived and studied log($N$)-log($S$) (Figure \ref{logN-logS}) and the unbinned XLF (Figure \ref{volume_density}).
The cluster X-ray source fraction is estimated to be $\sim$3$\sigma$ higher relative to the Lockman Hole at
$F_X$=1$\times$10$^{-14}$ ergs s$^{-1}$ flux limit.
These clusters were studied previously in optical wavelengths.
The galaxy luminosity function (GLF) is one of the basic statistics of the properties of galaxies.
Oemler (1974) studied 15 clusters' GLF including A194 and reported that the clusters and the blank-field have similar shape.
Christlein \& Zabludoff (2003) studied R-band GLFs of 6 nearby clusters, and A1060 was the nearest cluster in their survey sample.
Based on their study, a factor 75 higher galaxy density is estimated on average for the clusters with respect to non-cluster fields.
The highest fraction is measured at $M_R$ = $-$18.75 with $\Phi(M_R)/dM_R$ = 0.07 Mpc$^{-3}$ galaxies, 
which corresponds to a factor $\sim$175 above the non-cluster field density.
The luminosity range of our survey sample is 
2.88$\times$10$^{39}$ ergs s$^{-1}$ $\leq$ $L_X$ $\leq$ 1.15$\times$10$^{42}$  ergs s$^{-1}$.
At this low-luminosity levels emission from Low Mass X-Ray Binaries (LMXBs), 
hot halo and low-luminous AGNs (LLAGN) should be taken into consideration.
We investigate soft and hard emission by means of color-diagrams (Figure \ref{color-diagrams})
and found the majority of sources to have a hard spectral type.
If these galaxies do not possess an AGN,
the emission is expected to be produced by unresolved population of LMXBs (Sarazin et al. 2001; Blanton et al. 2001).
3 galaxies are detected with soft thermal emission from diffuse galactic gas.
The temperature values were found to be (0.2 keV $\leq$ kT $\leq$ 0.8 keV) 
in the average range depending the bulge size of the galaxy.

To address the origin of the X-ray emission we check (i) X-ray to optical luminosity ratio ($L_X/L_B$) 
and (ii) binned XLF by comparing with noncluster fields.
Since B-band magnitudes are sensitive to dust and recent star formation, 
X-ray to optical luminosity ratio ($L_X/L_B$) provide important constraints on ISM; hot halo and LMXBs 
(Matsushita 2001; Sarazin et al. 2001; Irwin et al. 2002).
Alternatively, if the sources without an optical counterparts are LLAGNs,
we can compare our XLF to AGN source density of groups of galaxies (Elvis et al. 1984) 
and that of the blank-fields (Miyaji et al. 2000; Ueda et al. 2003).

%

\begin{table*}
\centering
{\small
\begin{tabular}{l l c c c c c c c c c}     
\hline\hline
Src  & Name 		& $z$	& log ($r_{e}$)& $m_B$& $M_B$	 & log ($L_B/L_\odot$)	& log ($L_X$) & log ($L_X/L_B$)& Type & Group \\
(1)  & (2) 		& (3)	& (4) 		 & (5)   & (6)	 & (7)			& (8) 	      & (9)  		& (10)  & (11)  \\
\hline
\#1  & NGC 547 		& 0.018239& 0.951	& 12.91	& -21.42 &	10.73		& 42.25		& 31.52 & E1 & A194 \\
\#8  & NGC 541		& 0.018086& 1.417	& 13.07 	& -21.25 & 	10.66		& 40.83 		& 30.17 & S0 & A194 \\
\#23 & NGC 543		& 0.017666& 0.675	& 14.46	& -19.80 & 	10.09		& 39.08		& 28.99 & S0 & A194 \\
\#27 & [D80] 051		& 0.013219& 0.730	& 14.45	& -19.18 & 	 9.84			& 40.06		& 30.22 & S0 & A194 \\
\#56 & NGC 545		& 0.017806& 1.704	& 12.97	& -21.31 & 	10.69		& 41.10 		& 30.41 & S0 & A194 \\
\#1  & NGC 3312		& 0.009627& 1.421 	& 11.85 	& -21.10 & 	10.60 		& 40.94		& 30.34 & Sb & A1060 \\
\#3  & NGC 3311 		& 0.011985& 1.979	& 12.08 	& -21.34 & 	10.70		& 41.25		& 30.55 & E2 & A1060 \\
\#5  & NGC 3309		& 0.013593& 1.337 	& 12.12 	& -21.58 & 	10.79		& 40.51		& 29.72 & E3 & A1060 \\
\#11 & [S96a] 028		& 0.039604& 1.420 	& 12.78 	& -23.24 & 	11.46		& 41.28 		& 29.82 & SB & A1060 \\
\#12 & ESO501-G 047	& 0.016088& 1.937 	& 14.32	& -19.74 & 	10.06 		& 40.42		& 30.36 & S0 & A1060 \\
\#22 & NGC 3308		& 0.011855& 1.621 	& 12.78 	& -20.62 & 	10.41 		& 40.22		& 29.81 & S0 & A1060 \\	
\hline
\end{tabular}}
\caption{\label{lx_lb} Optical properties of galaxies identified for A1060 and A194. {\scriptsize NOTE-}
(1) The names given in this survey. (2) Galaxy names from {\scriptsize NASA} Extra-galactic Database (NED). 
The references for the names appearing in brackets are Stein 1996 [S96a] and Dressler 1984 [D80]. 
(3) redshift values from NED.
(4) Effective radius from RC3. 
(5) The values for the blue apparent magnitude, $m_B$, by RC3.
(6) Absolute blue magnitude, calculated from apparent magnitude and distance, as $M_B = m_B + 5 - 5 log(d)$.
(7) Blue-band luminosity in solar units, defined as log $L_B = -0.4 (M_B - 5.41)$.
(8) X-ray luminosities ($<$4$r_{e}$) of galaxies in ergs s$^{-1}$.
(9) Logarithmic X-ray to optical luminosity ratio.
(10) Morphological type code from RC3.
(11) Name of the group to which the galaxy belongs.
}
\end{table*}
%

   \begin{figure}
   {\centering
	\includegraphics[width=6.5cm,angle=270]{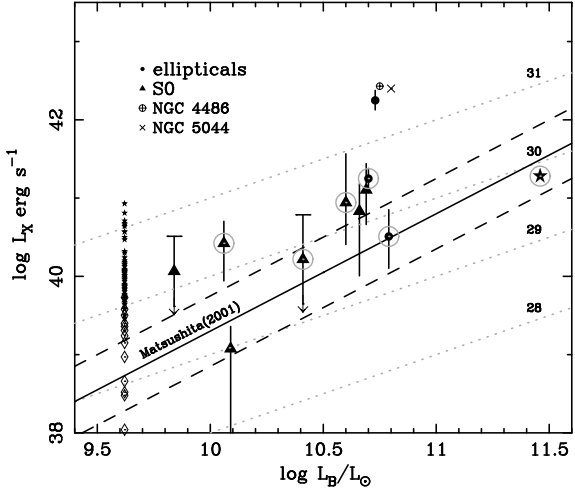}
    }
      \caption{\label{Lx_Lb}
$L_X$-$L_B$ relation for bright elliptical and spiral galaxies.
$L_X$ values are calculated within $<$4$r_e$ of galaxies in the 0.3$-$10 keV band, 
where $r_e$ is the effective radius for each galaxies.
Symbols indicate galaxy classifications for cluster elliptical ($filled$ $circles$),
cluster spirals ($filled$ $triangles$).
The circled sources are the ones from A1060 region. 
The $crossed$ $circle$ stands for NGC 4486 (Virgo cD) and the cross is the 
X-ray extended elliptical NGC 5044 from Matsushita (2001).
Gray dashed lines shows the slopes for log ($L_X/L_B$) ratios of 28, 29, 30 and 31.
Solid line is the distribution function for X-ray compact galaxies defined by Matsushita (2001)
with 90$\%$ confidence limits (dashed lines).}
   \end{figure}

\subsection{X-ray to optical luminosity relation}

The hard-band (2$-$10 keV) X-ray luminosity values are plotted against optical blue luminosities in Figure \ref{Lx_Lb}.
The properties of 11 optically identified galaxies are given in Table \ref{lx_lb}.
The effective radius of each galaxy, $r_{e}$, the distance from galaxy center 
within which half the optical luminosity is emitted,
and apparent blue magnitudes, $m_B$ are taken from de Vaucouleurs et al. (1991, hereafter RC3).
The absolute blue magnitudes, $M_B$ are calculated using $M_B$ = $m_B$+5$-$5log($d$), where $d$ is the distance in parsecs.
The blue-band luminosity values are determined using the relation log$L_B$ = $-$0.4($M_B$$-$5.41).
X-ray counts are accumulated within $r_x$=45 arcsec, which is the angular radius of 90\% encircled energy for PN.
X-ray emission can be considered compact, because $r_x$$\leq$4$r_e$ for the majority of the sources.
The dotted gray lines represents log($L_X/L_B$)= 28, 29, 30 and 31 for visual aid.
The straight line shows the average distribution of X-ray compact early-type galaxies and 
dashed line is the relative 90\% confidence limits reported by Matsushita (2001).
There is a clear scatter in Figure \ref{Lx_Lb}.
The most striking feature is that the total $L_X/L_B$ values of our sample are 
slightly higher than the compact early-type galaxies,
e.g., $L_X/L_B$=6.0$\times$10$^{29}$ ergs s$^{-1}$ $L^{-1}_{B\odot}$ for M81 (Sarazin et al. 2001). 
This is due either to a bright central AGNs or 
the contribution of more unresolved discrete sources than expected.
NGC 547 (from A194) is the most luminous X-ray source in our survey ($L_X$ =  10$^{42.25}$ ergs s$^{-1}$).
The $L_X/L_B$ relation of this elliptical galaxy resembles 
X-ray extended elliptical NGC 5044 and cD galaxy NGC 4486 (of Virgo) from Matsushita 2001 (Figure \ref{Lx_Lb}).
The entire X-ray emission radius, $r_x$, is defined to be the distance from the X-ray peak 
to the radius where the emission level is 1$\sigma$ higher than the surrounding background.
Based on the radial profile investigation of our survey galaxies, we calculate an average radius of $r_x$=15$^{\prime\prime}$.
The X-ray radius of NGC 547 has the highest value of $r_x$=19.8$^{\prime\prime}$, 
which can reasonably make the source an extended elliptical. 
NGC 547 is found to be much brighter than typical elliptical galaxies 
as expected for normal cD galaxies (Sandage \& Hardy 1973; Schombert 1986)
and located near the spatial and kinematical center of the host cluster A194.
We could not find any study in the literature defining NGC 547 as a cD galaxy.
There is a slight suspicion whether or not NGC 547 is the cD galaxy.
On the other hand, one of the brightest lenticular (S0) galaxy NGC 541 from A194 is reported to have an envelope by 
Andr\'{e}s J. et al. (2004) based on the $\emph{Hubble Space Telescope (HST)}$ and Very Large Telescope (VLT) observations. 
These findings are considered as an evidence that NGC 541 is the cD of A194.
From the $L_X/L_B$ ratio we do not get any peculiarity for NGC 541 (log $L_B$,$L_X$ = 10.66,40.83).

X-ray to optical luminosity ratios of the central galaxies of A1060 are also studied by Yamasaki et al. (2002) with $Chandra$.
They report $L_B$/$L_X$ ratios of 2.9 $\times$ 10$^{30}$ and 
1.3 $\times$ 10$^{30}$ ergs s$^{-1}$ $L^{-1}_{B\odot}$ for NGC 3311 and NGC 3309, respectively. 
The NGC 3311 $L_B$/$L_X$ ratio from our study is 1.3 times higher than the $Chandra$ result.
Whereas, NGC 3309 is a factor of 0.4 magnitude lower. 
Yamasaki et al. (2002) reports a possible underestimation of luminosities about a factor of 2 at maximum
as a result of the uncertainty of the quantum efficiency of the ACIS at low-energy band.
This can explain the result for NGC 3311.
The luminosity differences could also be due to the different sizes of the integration regions.
Other than that, the background integration region differs. 
We used annulus of 60$^{\prime\prime}$ $\leq$ R $\leq$ 90$^{\prime\prime}$ around the galaxies 
whereas a R$\leq$98$^{\prime\prime}$ is used for the analysis of the $Chandra$ data.
This may introduce a slight difference of the ICM contribution to the source spectrum.
Another possibility is the time variability of non-negligible LLAGNs luminosity in the center of these galaxies 
($Chandra$ on 2001 June and XMM on 2004 June).

The S0 galaxy NGC 543 (\#23 from A194 field) has the lowest $L_X/L_B$ value in our survey.
It has log$L_X/L_B$=28.99 which is a factor of 5.5 less than the nearby bulge-dominated Sa galaxy NGC 1291 (Irwin et al. 2002).
The X-ray to optical blue luminosity ratio of NGC 1291 is reported by Irwin et al. (2002) using $Chandra$ 
as factor of 1.4 less than the elliptical galaxy NGC 4697 (Sarazin et al. 2001) and the S0 galaxy NGC 1553 (Blanton et al. 2001).
This may be the result of not having luminous sources in NGC 543, as also stated for NGC 1291 by Irwin et al. (2002).
NGC 543 either has a faint LLAGN or there is no central active nuclei.
It probably has been striped its X-ray emitting gas by a recent interaction with other galaxies or the ICM. 
The source [D80] 051 (\#27 from A194) is a foreground source and 
has the lowest blue-luminosity value, $L_B$=0.69$\times$10$^{10}$ ergs s$^{-1}$ $L^{-1}_{B\odot}$.
The source [S96a] 028 (\#11) is a background SB from A1060 field with a $z$=0.0396 (circled star in Figure \ref{Lx_Lb}),
therefore we do not discuss this source with the other cluster members.

The $L_X/L_B$ plot (Figure \ref {Lx_Lb}) is obtained for the sources with optical counterparts,
which is less than 10\% of the X-ray sources.
We show the $L_X/L_B$ ratios for the other sources in the same figure.
The limiting magnitude is set to $M_B$ =  -18.64 ($L_B$=10$^{9.62}$ ergs s$^{-1}$ $L^{-1}_{B\odot}$) 
(private communication with H. Yamanoi).
Considering the error ranges of the X-ray luminosity and 90\% confidence range of 
the average $L_X/L_B$ distribution reported by Matsushita (2001), 
the X-ray emission from 13 more sources are found to be explained by unresolved LMXBs. 
To avoid visual confusion, we indicate their locations with diamonds (Figure \ref {Lx_Lb}) but do not present the error bars.
The remaining 52 sources, which is 68\% of the X-ray sources are not significantly contaminated by discrete sources.
In other words, LMXBs are insufficient to explain the observed X-ray emission from these sources.  
They are either optically obscured or powered by some other physical mechanism so that they become bright in X-rays.
The nature of X-Ray-bright optically-normal galaxies (XBONGs) are well studied by Comastri et al. (2002) and Yuan et al. (2004). 
One of the explanations is that XBONGs are luminous AGNs, which is also an ideal solution for our result.
We review the possibility of AGN-powered X-ray emission in the following section.

\subsection{\label{dlf} X-ray Luminosity Function}

\begin{figure}[ht] 
   {\centering
	\includegraphics[width=6.5cm,angle=-90]{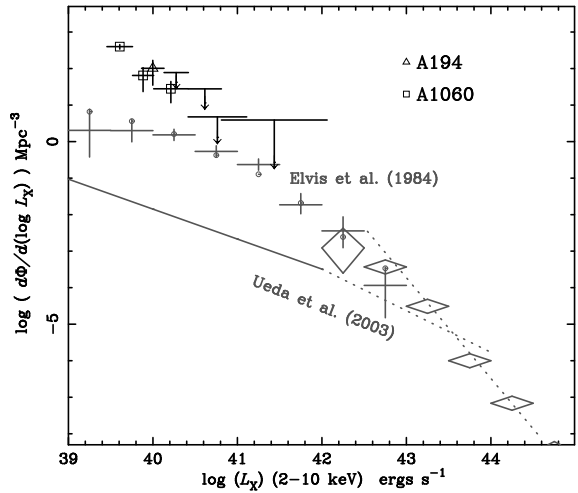}
     }
      \caption{ \label{Ueda_fig} 
Intrinsic X-ray luminosity function (XLF) of the sources in 2$-$10 keV energy band.
The black-crosses and solid-bars (stand for the upper-limits) are from the cluster fields.
The number density per unit volume per log $L_X$, $d\Phi(L_X)$/$d\log L_X$, versus $L_X$ in a logarithmically scaled plot.
The gray crosses show the predicted XLF from local group by Elvis et al. (1984). 
The curved dashed-line is the XLF for blank-fields, which is extrapolated from the result of Ueda et al. (2003)
for the analytical formula (Equation \ref{ueda.equ}) of Miyaji et al. (2000). (see $\S$ \ref{dlf})
	}\end{figure}

If the X-ray excess emission is assumed to be produced by something other than discrete sources, e.g. active nuclei, 
the intrinsic XLF can be comparable to non-cluster fields. 
In this study, we compare our cluster results to two different fields: groups of galaxies and blank skies. 
Figure \ref{Ueda_fig} shows the XLF computed for our survey.
The gray crosses show the predicted XLF from local group by Elvis et al. (1984). 
The diamonds show the luminosity function and the straight dashed line is the best-fit power law of $-$2.75 by Piccinotti el al. (1982).
As for the blank field, we employ the result of the cosmological evolution survey of Ueda et al. (2003), 
which is shown as the broken-dashed line.
Ueda et al. studied AGNs brighter than $L_X$=10$^{42}$ ergs s$^{-1}$.
In order to have the blank-field XLF for the lower luminosity levels of $L_X$ $\leq$ 10$^{42}$ ergs s$^{-1}$, 
we extrapolated the XLF reported by Ueda et al..
The extrapolation is estimated by an analytical expression of smoothly connected two power-law form as defined by Miyaji et al (2000):

\begin{equation}
  \frac{d\Phi(L_X,z=0)}{d \log L_X} = A \left [ \left (\frac{L_x}{L_{\ast}} \right )^{\gamma 1}+ \left (\frac{L_x}{L_{\ast}} \right )^{\gamma 2} \right ]^{-1}  
\label{ueda.equ}
\end{equation}

The $d\Phi(L_X,z)/\log L_X$ represents the volume density per $\log L_X$ as a function of $L_X$ and $z$.
Using the best-fit parameters of the pure luminosity evolution (PLE) model from Ueda et al. (2003), 
($A = 1.41 \times 10^{-6}$ Mpc$^{-3}$, $\log L_{\ast} = 43.66$ ergs s$^{-1}$, ${\gamma 1} =$ 0.82, ${\gamma 2} =$ 2.37)
we calculated luminosity function for local universe ($z\leq$0.02).
The black crosses at the top of Figure \ref{Ueda_fig} represent the computed XLF for the clusters.
Our X-ray selected sources have luminosity values between 10$^{39.6}$$\leq$$L_X$$\leq$10$^{41.4}$ ergs s$^{-1}$.
The luminosity restrictions are introduced by the instrumental incompleteness for faint sources and statistical shortage for bright sources.
The source density decreases gradually for the sources brighter than $L_X$$\geq$10$^{40.5}$ ergs s$^{-1}$, 
at which point the source density is comparable to the field level.
A source density excess is evident at low luminosity levels of $L_X$$\leq$10$^{40.5}$ ergs s$^{-1}$.
The source density for the clusters is found to be around 3 orders of magnitude higher than the field.
Christlein \& Zabludoff (2003) report cluster source density as high as a factor 175 larger than the blank-fields in optical bands.
An equal amount of overdensity in X-rays would be the expected level.
There is a minimum of a factor 6 higher source density observed in X-rays compared to the blank field.
The excess is a factor of 15 higher relative to the source density of the local group, reported by Elvis et al. (1984).
A comparison of the cluster source density with both non-cluster fields implies an elevated AGN activity in the cluster environment. 

It is known that galaxies funnel toward clusters through filaments, where galaxies appear like pearls on a string. 
The projection effect may cause an enhancement of the source density, if a filament aligns through the line of sight. 
Simulations of the large scale formations (Gavernato et al. 1998) show a quantitatively 
higher number of densities on knots (clusters) than filaments.  
If one assumes that the observed X-ray excess is due to the projectional effects, 
the required length of a filament with 1 Mpc average radius would be about 100 Mpc. 
Since mean cluster separation is several 10 Mpc, 
the projection effect of the filamentary branches itself would be unrealistic. 
However, a partial enhancement from filaments is admitted. 

Recent ${\chandra}$ studies show that X-ray sources are not distributed randomly (Martini et al. 2002; Johnson et al. 2003).
Bright sources are reported to populate at the outskirts, whereas faint ones are found somewhat nearer to the cluster center ($<$ 1 Mpc).
Based on the radial distribution of the sources, the possibilities of AGN fueling and quenching are suggested by the ${\chandra}$ results. 
Considering the restricted low luminosity range (10$^{39.6}$$\leq$$L_X$$\leq$10$^{41.4}$ ergs s$^{-1}$)
and X-ray to optical luminosity ratios ($L_X/L_B$) from our survey,
the X-ray sources from the fields of A194 and A1060 are likely to be experiencing a similar physical situation. 
The fueling scenario can account for the excess for the faint sources and 
the quenching can explain the population vanishing for the bright sources as obtained from the XLF (Figure \ref{Ueda_fig}).
The dense cluster environments are the places with high possibility of collisions and 
close encounters of galaxies in the vicinity of clusters.
When a galaxy falls into the cluster gravitational potential, the balance between AGN and surrounding gas is disturbed.
The gas falls on to AGN and powers it.
The infall induces AGN activity, therefore the source gets brighter. 
It is currently believed that most -if not all- galaxies, including our galaxy, host a Black Hole (BH) at its center. 
Most of them are inactive not accreting much matter. 
The physical conditions in the ICM may also awake these kinds of starved BHs, which increases the X-ray sources density in clusters.
However,  the mechanism reverses at smaller distances from the center as the galaxy loses a significant amount of its fuel by ram-pressure striping.
AGN activity is reduced by deep cluster potential.
In other words, AGN activity decreases for the longer residents.
This explains why the source density vanishes at luminosity levels $L_X$ $\geq$ 10$^{40.5}$ ergs s$^{-1}$
and the absence of the bright sources $L_X$ $\geq$ 10$^{42}$ ergs s$^{-1}$ in nearby poor clusters.

\section{Conclusions}

We have presented the nature of X-ray detected sources in the fields of nearby, poor clusters of galaxies A194 and A1060.
A total of 76 X-ray sources have been detected (46 and 32, respectively) by the PN camera. 
X-ray spectrum of several sources are studied for the first time, particularly for the A194 sources. 
The sources with low counting statistics have been studied using the so called X-ray color-diagrams. 
The best-fit values of used plasma models
are found to be in the reasonable range, the temperatures are sub keV levels (kT$\leq$1.0 keV)
and photon indices are around $\Gamma$=1.7$\pm{1.0}$.
The excess of the X-ray sources at $F_X$=1$\times$10$^{-14}$ ergs cm$^{-2}$ s$^{-1}$ is 
estimated as $\sim$3$\sigma$ from the log($N$)-log($S$) plot.
Considering low luminosity levels of our survey sample of $L_X$$\leq$10$^{42}$ ergs s$^{-1}$, 
the fractions of the X-ray emission produced by LMXB, hot halo and LLAGN are evaluated separately.
Based on the $L_X/L_B$ relation, about 30\% of the whole sample can be explained by the X-ray emission from unresolved LMXBs.
Since the spectroscopic redshift is available for 10\% of the X-ray sources,
the cluster source density is approximated by subtracting the field source density.
The difference is attributed to galaxies in the cluster gravitational potential. 
The X-ray luminosity function exhibits a source density excess for the clusters by a factor of 6 higher than the blank field 
and a factor of 15 higher than the local group at $L_X$$\geq$10$^{40.5}$ ergs s$^{-1}$.
The enhancement vanishes for the brighter sources.
We suggest that AGN fueling is induced and inactive BHs are awaken in the outskirts, 
whereas AGN activity is quenched in the cluster center.  

We would like to thank Chiho Matsumoto and Yoshihiro Ueda, for valuable discussions and 
Hitomi Yamanoi, for providing the SUBARU observations of Abell 1060 before its publication.
This work is supported in part by a Grant-in-Aid for Scientific Research on Specially Promoted 
Research, contract No.15001002, from the Ministry of Education, Sports, Culture, Science and Technology (MEXT), Japan.
The author also acknowledges the support for overseas research students by MEXT.
This research has made use of data obtained through the $XMM$-$Newton$, an ESA Science Mission with instruments
and contributions directly funded by ESA Member states and the US (NASA).


\appendix{APPENDIX}
The detected sources are listed in Table \ref{table_A194} and Table \ref{table_A1060} for A194 and A1060.
The columns are; (1) source number in this survey. (2) acronym with RA-Dec. (3) positional errors in arcsec. 
(4), (5), (6) The source counts in ks for soft, medium and hard band. (7) Hardness ratio 1, HR1 = (M-S/M+S).
(8) Hardness ratio 2, HR2 = (H-M)/(H+M). (9) Flux in 2-10 keV. (10) Luminosity in 2-10 keV. 
(11) Counterpart in other bands (12) optical redshift.

\begin{sidewaystable*}
\begin{minipage}[t][180mm]{\textwidth}
\centering          
{\scriptsize
\caption{\label{table_A194}      
A194 - EPN source counts and HR values}
\begin{tabular}{ c c c c c c c c c c c c c }
\hline
SRC & ACRONYM & R$\sigma$ & \multicolumn{3}{ c }{Count rate (count ks$^{-1}$)} &  HR1 &  HR2 & $F_{[2-10]}\times$10$^{-14}$ & Log $L_X$ & Note & $z_{opt}$\\ 
    & RA Dec  & arcsec & SOFT & MID & HARD &  (M-S/M+S)   & (H-M)/(H+M) &   ergs cm$^{-2}$ s$^{-1}$ & & & \\
 (1)   & (2)  & (3) & (4) & (5)  & (6)  &  (7)   & (8) &   (9)  & (10) & (11) & (12) \\
\hline
\hline
1 & XMMU J012600.5-012043 & 0.20 & 84.89${\pm}$3.30 & 19.70${\pm}$1.64 & 42.91${\pm}$2.39 & -0.62${\pm}$0.03 & 0.37${\pm}$0.04 	& 159.13 & 42.06 & NGC 547 & 0.01823 \\
2 & XMMU J012535.9-012546 & 0.41 & 25.06${\pm}$1.64 & 10.80${\pm}$1.09 & - & -0.40${\pm}$0.05 & -1.00${\pm}$0.00 		& 5.97 & 40.64 & & \\
3 & XMMU J012549.3-012412 & 0.54 & 7.94${\pm}$0.95 & 3.72${\pm}$0.69 & 4.47${\pm}$0.77 & -0.36${\pm}$0.10 & 0.09${\pm}$0.13 	& 1.97 & 40.15 & & \\
4 & XMMU J012602.8-012701 & 0.82 & - & 2.65${\pm}$0.59 & 9.28${\pm}$1.13 & 1.00${\pm}$0.00 & 0.56${\pm}$0.09			& 7.05 & 40.71 & & \\
5 & XMMU J012529.6-011455 & 0.81 & 4.32${\pm}$0.92 & 5.33${\pm}$0.92 & 8.82${\pm}$1.54 & 0.10${\pm}$0.14 & 0.25${\pm}$0.12 	& 6.61 & 40.68 & & \\
8 & XMMU J012544.2-012247 & 0.51 & 35.35${\pm}$3.88 & 4.22${\pm}$1.16 & 8.67${\pm}$1.14 & -0.79${\pm}$0.06 & 0.34${\pm}$0.13 	& 3.62 & 40.42 & NGC 541 & 0.01808 \\
9 & XMMU J012556.3-012515 & 0.57 & 10.62${\pm}$1.08 & 5.46${\pm}$0.76 & 3.10${\pm}$0.73 & -0.32${\pm}$0.08 & -0.28${\pm}$0.13 	& 1.66 & 40.08 & & \\
10 & XMMU J012611.0-012510 & 1.34 & - & 2.00${\pm}$0.55 & - & 1.00${\pm}$0.00 & -1.00${\pm}$0.00 				& 0.52 & 39.58 & & \\
11 & XMMU J012540.1-011945 & 0.61 & 8.15${\pm}$1.03 & 3.72${\pm}$0.69 & 3.68${\pm}$0.76 & -0.37${\pm}$0.10 & -0.01${\pm}$0.14 	& 1.41 & 40.01 & & \\
12 & XMMU J012554.5-012343 & 0.57 & 10.46${\pm}$1.05 & 3.27${\pm}$0.62 & - & -0.52${\pm}$0.08 & -1.00${\pm}$0.00 		& 1.97 & 40.16 & & \\
13 & XMMU J012555.5-011119 & 1.00 & 6.10${\pm}$1.34 & 5.42${\pm}$1.17 & - & -0.06${\pm}$0.15 & -1.00${\pm}$0.00 		& 4.04 & 40.47 & & \\
14 & XMMU J012533.1-013333 & 1.73 & 4.53${\pm}$0.98 & - & - & -1.00${\pm}$0.00 & -						& 1.52 & 40.04 & & \\
16 & XMMU J012541.3-012836 & 1.34 & - & - & 2.55${\pm}$0.62 & - & 1.00${\pm}$0.00 						& 0.20 & 39.15 & & \\
17 & XMMU J012554.5-012517 & 1.36 & - & - & 2.90${\pm}$0.69 & -1.00${\pm}$0.00 & 1.00${\pm}$0.00 				& 0.98 & 39.85 & & \\
18 & XMMU J012538.7-013107 & 1.22 & 2.45${\pm}$0.65 & - & 3.50${\pm}$0.84 & - & 1.00${\pm}$0.00 				& 0.40 & 39.46 & & \\
19 & XMMU J012622.5-012654 & 1.34 & - & - & 6.66${\pm}$1.30 & - & 1.00${\pm}$0.00 						& 2.75 & 40.30 & & \\
20 & XMMU J012614.9-012240 & 1.44 & - & - & 3.49${\pm}$0.82 & - & 1.00${\pm}$0.00						& 0.75 & 39.73 & & \\
21 & XMMU J012544.0-011728 & 0.70 & 7.11${\pm}$1.01 & 3.58${\pm}$0.67 & 3.19${\pm}$0.76 & -0.33${\pm}$0.11 & -0.06${\pm}$0.15 	& 1.09 & 39.90 & & \\
22 & XMMU J012635.5-012349 & 1.56 & 3.81${\pm}$0.98 & 2.65${\pm}$0.76 & - & -0.18${\pm}$0.19 & -1.00${\pm}$0.00 		& 0.53 & 39.58 & & \\
24 & XMMU J012551.7-012919 & 0.97 & 5.78${\pm}$0.92 & 2.84${\pm}$0.60 & 2.54${\pm}$0.63 & -0.34${\pm}$0.12 & -0.06${\pm}$0.16 	& 1.25 & 39.96 & & \\
25 & XMMU J012612.0-012904 & 0.94 & 4.79${\pm}$0.88 & 2.68${\pm}$0.61 & 5.17${\pm}$1.09 & -0.28${\pm}$0.13 & 0.32${\pm}$0.14 	& 1.87 & 40.13 & & \\
28 & XMMU J012556.5-012732 & 0.92 & 4.54${\pm}$0.78 & - & - & -1.00${\pm}$0.00 & - 						& 0.13 & 38.97 & \\
29 & XMMU J012503.2-012244 & 0.81 & 8.32${\pm}$1.32 & 5.20${\pm}$0.97 & 10.20${\pm}$1.68 & -0.23${\pm}$0.12 & 0.32${\pm}$0.11 	& 6.36 & 40.66 & & \\
30 & XMMU J012611.2-012802 & 1.42 & 3.48${\pm}$0.75 & - & - & -1.00${\pm}$0.00 & - 						& 1.19 & 39.94 & & \\
31 & XMMU J012531.9-012232 & 0.95 & 4.19${\pm}$0.76 & 2.44${\pm}$0.59 & - & -0.26${\pm}$0.14 & -1.00${\pm}$0.00 		& 1.42 & 40.01 & & \\
32 & XMMU J012539.3-011309 & 1.05 & 6.86${\pm}$1.51 & - & 6.73${\pm}$1.83 & -1.00${\pm}$0.00 & 1.00${\pm}$0.00 			& 5.11 & 40.57  & & \\
33 & XMMU J012539.1-012126 & 0.90 & 4.13${\pm}$0.78 & 2.26${\pm}$0.54 & 2.79${\pm}$0.67 & -0.29${\pm}$0.14 & 0.10${\pm}$0.17 	& 1.06 & 39.88 & & \\
35 & XMMU J012600.5-012747 & 1.19 & 3.12${\pm}$0.71 & - & 2.74${\pm}$0.69 & -1.00${\pm}$0.00 & 1.00${\pm}$0.00 			& 2.18 & 40.20 & & \\
36 & XMMU J012612.2-012953 & 2.16 & 6.15${\pm}$1.64 & - & - & -1.00${\pm}$0.00 & - 						& 11.87 & 40.93 & & \\
37 & XMMU J012557.9-011717 & 0.89 & 7.95${\pm}$1.13 & 3.21${\pm}$0.67 & 3.91${\pm}$0.87 & -0.43${\pm}$0.10 & 0.10${\pm}$0.15 	& 0.88 & 39.81 & & \\
38 & XMMU J012611.1-011155 & 0.61 & 28.23${\pm}$2.69 & 10.10${\pm}$1.59 & 6.44${\pm}$1.71 & -0.47${\pm}$0.07 & -0.22${\pm}$0.15 & 0.70 & 39.71 & & \\
41 & XMMU J012610.6-012349 & 1.59 & - & - & 4.42${\pm}$0.89 & - & 1.00${\pm}$0.00 						& 2.42 & 40.24 & & \\
43 & XMMU J012510.7-011542 & 0.93 & 13.65${\pm}$1.74 & 5.58${\pm}$1.27 & - & -0.42${\pm}$0.11 & -1.00${\pm}$0.00 		& 2.85 & 40.32 & & \\
44 & XMMU J012624.6-012415 & 1.76 & 2.76${\pm}$0.72 & - & - & -1.00${\pm}$0.00 & - 						& 0.81 & 39.77 & & \\
45 & XMMU J012513.0-012802 & 1.27 & 3.33${\pm}$0.78 & 2.18${\pm}$0.64 & 5.06${\pm}$1.12 & -0.21${\pm}$0.18 & 0.40${\pm}$0.15	& 1.01 & 39.87 & RadioS. & \\
46 & XMMU J012556.6-012610 & 2.62 & - & - & 2.97${\pm}$0.70 & - & 1.00${\pm}$0.00 						& 8.92 & 40.81 & & \\
47 & XMMU J012554.9-012927 & 1.46 & 2.73${\pm}$0.68 & 2.17${\pm}$0.53 & 3.05${\pm}$0.79 & -0.11${\pm}$0.17 & 0.17${\pm}$0.17 	& 5.10 & 40.57 & & \\
48 & XMMU J012508.2-011815 & 1.81 & - & 2.81${\pm}$0.80 & - & 1.00${\pm}$0.00 & -1.00${\pm}$0.00 				& 0.54 & 39.59 & & \\
49 & XMMU J012528.6-012430 & 1.93 & 4.47${\pm}$0.94 & - & - & -1.00${\pm}$0.00 & - 						& 0.02 & 38.04 & & \\
50 & XMMU J012534.3-011640 & 5.23 & 3.40${\pm}$0.84 & - & - & -1.00${\pm}$0.00 & - 						& 1.61 & 40.07 & & \\
51 & XMMU J012601.9-012808 & 3.26 & 2.32${\pm}$0.60 & - & - & -1.00${\pm}$0.00 & - 						& 4.33 & 40.50 & & \\
52 & XMMU J012606.1-011744 & 3.41 & 3.41${\pm}$0.83 & - & - & -1.00${\pm}$0.00 & - 						& 0.35 & 39.40 & & \\
53 & XMMU J012627.9-011808 & 3.92 & 4.15${\pm}$1.23 & - & - & -1.00${\pm}$0.00 & - 						& 0.28 & 39.31 & & \\
54 & XMMU J012544.0-012732 & 2.04 & 2.97${\pm}$0.70 & - & - & -1.00${\pm}$0.00 & - 						&  -   & -     & & \\
55 & XMMU J012623.7-011733 & 2.42 & 5.15${\pm}$1.29 & - & - & -1.00${\pm}$0.00 & - 						& 0.65 & 39.67 & & \\
56 & XMMU J012559.0-012027 & ...   & 26.72${\pm}$1.84 & 11.29${\pm}$1.22 & 10.19${\pm}$1.33 & -0.05${\pm}$0.08 & -0.41${\pm}$0.12 & 10.77 & 40.89 & NGC 545 & 0.01780 \\
\hline
\end{tabular}}
\end{minipage}
\end{sidewaystable*}

\begin{sidewaystable*}
\begin{minipage}[t][180mm]{\textwidth}
\centering          
{\scriptsize
\caption{\label{table_A1060} A1060 - EPN source counts and HR values. $\ast$: NGC 3314 is a galaxy pair, having $z =$00951 (NGC 3314A) and $z =$01548 (NGC 3314B).}
\begin{tabular}{ c c c c c c c c c c c c c }
\hline
SRC & ACRONYM & R$\sigma$ & \multicolumn{3}{ c }{Count rate (count ks$^{-1}$)} &  HR1 &  HR2 & $F_{[2-10]}\times$10$^{-14}$ & Log $L_X$ & Note & $z_opt$\\ 
    &  & arcsec & SOFT & MEDIUM & HARD &  (M-S/M+S)   & (H-M)/(H+M) &   ergs cm$^{-2}$ s$^{-1}$ & ergs s$^{-1}$ & & \\
\hline
\hline
1 & XMMU J103702.7-273354.5 & 0.22 & 23.23${\pm}$1.42 & 6.60${\pm}$0.83 & 39.74${\pm}$1.83 & -0.56${\pm}$0.05  	& 0.72${\pm}$0.03 	& 51.05 & 41.12 & NGC 3312 & 0.00962 \\
2 & XMMU J103616.2-273945.4 & 0.44 & - & 4.01${\pm}$0.84 & 23.03${\pm}$1.79 		& 1.00${\pm}$0.00 	& 0.70${\pm}$0.06 	& 10.32 & 40.42 & \\
3 & XMMU J103643.0-273142.8 & 0.27 & 40.40${\pm}$1.78 & 23.04${\pm}$1.38 & 18.08${\pm}$1.44 & -0.27${\pm}$0.03 	& -0.12${\pm}$0.05 	& 23.35 & 40.78 & NGC 3311 & 0.01198 \\
4 & XMMU J103626.9-273819.6 & 0.32 & 24.56${\pm}$1.54 & 9.59${\pm}$1.03 & 13.78${\pm}$1.31 & -0.44${\pm}$0.05 	& 0.18${\pm}$0.07 	&  5.98 & 40.18 & \\
5 & XMMU J103635.9-273106.9 & 0.27 & 41.19${\pm}$1.84 & 16.59${\pm}$1.28 & 8.01${\pm}$1.14 & -0.43${\pm}$0.04 	& -0.35${\pm}$0.07 	& 1.39  & 39.55 & NGC 3309 & 0.01359\\
6 & XMMU J103715.0-272532.5 & 0.36 & 21.57${\pm}$1.43 & 7.84${\pm}$0.91 & 9.04${\pm}$1.12 & -0.47${\pm}$0.05 	& 0.07${\pm}$0.08 	& 2.16  & 39.74 & \\
7 & XMMU J103625.9-273729.5 & 0.38 & 15.04${\pm}$1.29 & 6.35${\pm}$0.88 & 11.97${\pm}$1.24 & -0.41${\pm}$0.07 	& 0.31${\pm}$0.08 	& 4.51 & 40.06 & \\
8 & XMMU J103645.6-272559.2 & 0.46 & 8.36${\pm}$0.96 & 5.44${\pm}$0.76 & 6.67${\pm}$0.89 & -0.21${\pm}$0.09 	& 0.10${\pm}$0.10 	& 2.05 & 39.72 & \\
9 & XMMU J103658.4-272057.1 & 0.57 & 10.14${\pm}$1.17 & 5.87${\pm}$0.85 & 5.93${\pm}$1.10 & -0.27${\pm}$0.09 	& 0.01${\pm}$0.12 	& 2.88 & 39.87 & \\
10 & XMMU J103659.8-273028.6 & 0.41 & 13.04${\pm}$1.08 & 6.86${\pm}$0.82 & 6.20${\pm}$0.88 & -0.31${\pm}$0.07 	& -0.05${\pm}$0.09 	& 3.03 & 39.89 & \\
11 & XMMU J103628.6-272237.2 & 0.48 & 9.64${\pm}$1.11 & 4.46${\pm}$0.78 & 11.07${\pm}$1.25 & -0.37${\pm}$0.09 	& 0.43${\pm}$0.09 	& 3.59 & 39.96 & [S96a] 028 & 0.03960 \\
12 & XMMU J103717.2-272809.1 & 0.78 & 4.20${\pm}$0.79 & 2.39${\pm}$0.59 & 6.60${\pm}$0.94 & -0.27${\pm}$0.14 	& 0.47${\pm}$0.11	& 4.04 & 40.01 & ESO501-G47 & 0.01608 \\
13 & XMMU J103714.2-273306.6 & 0.84 & 3.93${\pm}$0.84 & 4.05${\pm}$0.87 & 4.79${\pm}$0.96 & 0.02${\pm}$0.15 	& 0.08${\pm}$0.15 	& 2.71 & 39.84 & \\
14 & XMMU J103638.7-274250.7 & 0.89 & - & 4.41${\pm}$0.81 & 7.59${\pm}$1.24 & 1.00${\pm}$0.00 			& 0.26${\pm}$0.11 	& 4.77 & 40.09 & \\
15 & XMMU J103643.3-271927.5 & 1.32 & - & - & 5.42${\pm}$1.19 & - 						& 1.00${\pm}$0.00 	& 3.81 & 39.99 & \\
16 & XMMU J103712.8-274102.8 & 1.21 & 8.22${\pm}$1.10 & - & - 				& -1.00${\pm}$0.00 	& - 			& 4.63 & 40.07 & NGC 3314 & $\ast$ \\
17 & XMMU J103610.4-272850.4 & 1.13 & 5.35${\pm}$0.93 & - & 3.84${\pm}$0.93 & -1.00${\pm}$0.00 			& 1.00${\pm}$0.00 	& 1.25 & 39.50 &  &  \\
18 & XMMU J103620.0-273927.3 & 1.14 & 6.94${\pm}$1.13 & - & - & -1.00${\pm}$0.00 				& - 			& - & - & \\
19 & XMMU J103729.1-274234.7 & 1.34 & - & 4.73${\pm}$0.95 & - & 1.00${\pm}$0.00 				& -1.00${\pm}$0.00 	& 4.36 & 40.05 & \\
20 & XMMU J103638.6-271843.8 & 0.98 & 5.50${\pm}$1.01 & - & 6.04${\pm}$1.25 & -1.00${\pm}$0.00 			& 1.00${\pm}$0.00 	& 0.18 & 38.66 & \\
21 & XMMU J103748.7-272959.7 & 1.10 & 10.48${\pm}$1.33 & 3.14${\pm}$0.79 & - & -0.54${\pm}$0.10 		& -1.00${\pm}$0.00 	& 2.27 & 39.76 & \\
22 & XMMU J103622.5-272618.9 & 0.77 & 7.97${\pm}$1.04 & - & - & -1.00${\pm}$0.00 				& - 			& 0.13 & 38.5 & NGC 3308 & 0.01185 \\
23 & XMMU J103629.9-272133.4 & 1.05 & 3.35${\pm}$0.80 & 2.62${\pm}$0.70 & - & -0.12${\pm}$0.18 			& -1.00${\pm}$0.00 	& 0.12 & 38.48 & \\
24 & XMMU J103603.5-272958.6 & 1.06 & 5.58${\pm}$0.98 & - & - & -1.00${\pm}$0.00 				& - 			& 1.72 & 39.64 & \\
25 & XMMU J103704.4-272358.1 & 1.39 & 3.37${\pm}$0.90 & - & - & -1.00${\pm}$0.00 				& - 			& 1.48 & 39.58 & [S96a] 057 &  0.00916 \\
26 & XMMU J103709.3-273513.2 & 0.82 & 6.31${\pm}$0.83 & 2.75${\pm}$0.60 & - & -0.39${\pm}$0.11 			& -1.00${\pm}$0.00 	& 1.73 & 39.65 & \\
27 & XMMU J103558.0-273002.4 & 1.52 & 4.53${\pm}$0.92 & - & - & -1.00${\pm}$0.00 				& - 			& 0.90 & 39.36 & \\
28 & XMMU J103737.1-273601.5 & 1.74 & 5.21${\pm}$1.29 & - & 16.29${\pm}$3.86 & -1.00${\pm}$0.00 		& 1.00${\pm}$0.00 	& 2.64 & 39.83 & \\
29 & XMMU J103650.7-273549.6 & 1.27 & 3.51${\pm}$0.81 & 3.71${\pm}$0.72 & 3.71${\pm}$0.82 & 0.03${\pm}$0.15 	& 0.00${\pm}$0.15 	& 0.68 & 39.24 & \\
30 & XMMU J103645.4-274157.2 & 1.01 & 7.62${\pm}$1.02 & - & 3.73${\pm}$0.94 & -1.00${\pm}$0.00 			& 1.00${\pm}$0.00 	& 1.31 & 39.46 & \\
31 & XMMU J103737.5-272900.8 & 2.37 & - & 2.41${\pm}$0.65 & - & 1.00${\pm}$0.00 				& -1.00${\pm}$0.00 	& 4.52 & 40.06 & \\
32 & XMMU J103634.8-274159.8 & 1.42 & 3.75${\pm}$0.86 & 2.69${\pm}$0.71 & - & -0.16${\pm}$0.17 			& -1.00${\pm}$0.00 	& 3.67 & 39.97 & \\
\hline
\end{tabular}}
\end{minipage}
\end{sidewaystable*}

\end{document}